\documentstyle[11pt,aaspp4]{article}
\received{15 April 1996}
\input tex.def
\slugcomment{To appear in {\it The Astrophysical Journal (Letters)}}

\begin{document}

\title{Dynamical Evidence for a Luminous, Young Globular Cluster in
NGC 1569 \footnote{Based on observations obtained at the W. M. Keck 
Observatory.}}

\author{Luis C. Ho}
\affil{Harvard-Smithsonian Center for Astrophysics, Cambridge, MA 02138}

\and

\author{Alexei V. Filippenko}
\affil{Department of Astronomy, University of California, Berkeley, CA 94720-3411}

\begin{abstract}
Recent high-resolution observations with the {\it Hubble Space Telescope (HST)}
reveal that star clusters of extraordinary luminosity and compactness are
commonly found in a variety of starburst systems.  There has been
much speculation that these clusters represent present-day analogs of
young globular clusters.  Using the HIRES echelle spectrograph on the Keck 
10~m telescope, we obtained high-dispersion optical spectra of one of the 
``super star clusters'' (cluster ``A'') in the nearby dwarf galaxy NGC~1569.  
The size of the cluster is known from published {\it HST} images.  The 
line-of-sight velocity dispersion ($\sigma_*$ = 
15.7$\pm$1.5 \kms) has been measured from a cross-correlation analysis of its 
integrated spectrum at visual wavelengths.  If the cluster is gravitationally 
bound and the velocities are isotropic, application of the virial theorem 
implies that the cluster has a total stellar mass of (3.3$\pm$0.5)\e{5} 
\solmass.  This object's mass, mass density, and probable mass-to-light ratio 
after aging 10--15 Gyr are fully consistent with the typical values 
of Galactic globular clusters.  Our result strongly suggests that at 
least some of the luminous, compact, young star clusters being discovered with 
{\it HST} will indeed evolve into normal globular clusters of the type seen 
in the Milky Way.
\end{abstract}

\keywords{galaxies: individual (NGC 1569) --- galaxies: 
irregular --- galaxies: starburst --- galaxies: star clusters --- globular 
clusters: general}

\section{Introduction}
In the last few years, {\it Hubble Space Telescope (HST)} imaging studies of a 
variety of extragalactic star-forming systems have identified a widespread
new class of star clusters.  The compactness and high luminosities of these 
objects, coupled with their inferred youth, have stimulated speculation that 
they represent present-day analogs of young globular clusters.  Although the 
existence of a few such ``super star clusters'' 
had been known from previous ground-based studies (e.g., Arp \& Sandage 
1985; Melnick, Moles, \& Terlevich 1985; Lutz 1991), it took the resolving 
power of {\it HST} to demonstrate the prevalence of this phenomenon.  Such 
clusters appear to be a common feature in amorphous galaxies (O'Connell, 
Gallagher, \& Hunter 1994; Hunter, O'Connell, \& Gallagher 1994; O'Connell 
\etal 1995), merging and interacting systems (Holtzman \etal 1992; 
Whitmore \etal 1993; Whitmore \& Schweizer 1995), circumnuclear star-forming 
rings (Benedict \etal 1994; Barth \etal 1995; Bower \& Wilson 1995; 
Maoz \etal 1996), and various other starburst systems (Conti \& Vacca 
1994; Vacca 1994; Meurer \etal 1995).  

That globular clusters may be forming in the present epoch is an exciting 
development, as observational studies of such a process are relevant to issues 
ranging from large-scale star formation to galaxy formation and evolution.  
The hypothesis that the recently discovered clusters are young {\it globular} 
clusters rests on three pieces of evidence.  First, these clusters are 
generally quite compact, being unresolved or marginally resolved in {\it HST} 
images of relatively nearby galaxies, which implies that they have half-light 
radii of only a few parsecs, similar to the sizes of Galactic globular 
clusters.  Second, the brightest members have rather extraordinary 
luminosities, in many cases surpassing that of the R136 cluster in the center 
of the 30 Doradus complex in the Large Magellanic Cloud.  Some of the clusters 
in the studies mentioned above, for instance, have absolute visual magnitudes 
exceeding --14 to --15; for comparison, $M_V$ = --11.3 mag for R136 (O'Connell 
\etal 1994).  Finally, the blue optical continuum colors, and, where available, 
the amount of ultraviolet radiation, indicate ages ranging from a 
few to 500 Myr.  The presence of young stars has been confirmed unambiguously 
in several cases where spectroscopy with ground-based facilities (Arp \& 
Sandage 1985; Melnick \etal 1985; Schweizer \& Seitzer 1993; Prada, Greve, \& 
McKeith 1994; Zepf \etal 1995) or {\it HST} (Leitherer \etal 1996; Conti, 
Leitherer, \& Vacca 1996) has been feasible.  Population synthesis models 
generally indicate that the luminosities of these clusters will fade to the 
observed luminosities of old globular clusters in 10--15 Gyr, provided 
that they are bound and dynamical evolution does not dissolve them.  The 
shortness of the expected time scale for expansion, as deduced from the small 
physical dimensions, relative to their ages can be taken as evidence that 
most of the clusters may in fact be gravitationally bound (Whitmore \etal 
1993; Whitmore \& Schweizer 1995; Maoz \etal 1996).  The masses derived from 
the models lie in the range found in Galactic globular clusters.

The arguments in favor of the globular cluster interpretation would be 
considerably strengthened by a {\it direct}, model-independent measurement of 
the cluster masses.  Moreover, by combining the mass and the present 
luminosity with a model prediction of the luminosity evolution, one can 
compare the future mass-to-light ratio of the cluster to 
that of evolved globular clusters.  Such a comparison can put meaningful 
constraints on the stellar population, and hence the initial mass function, 
of the young clusters.  

This {\it Letter} reports the first 
attempt to measure the dynamical mass of one of these star clusters.  The
observations consist of high-dispersion (echelle) spectra of one of the 
two luminous star clusters in the central region of the nearby dwarf galaxy 
NGC~1569, which has recently been studied by O'Connell \etal (1994) 
using {\it HST}.  For an overview of the general properties of the galaxy and 
its two bright clusters, consult Arp \& Sandage (1985), 
Israel (1988), and O'Connell \etal (1994).  The line-of-sight stellar velocity 
dispersion is combined with the previously reported size measurement to 
estimate the dynamical mass.  The derived cluster mass ($\sim$3.3\e{5} 
\solmass) falls slightly higher than the peak of the mass function of evolved 
globular clusters in the Milky Way.  Here, we restrict our attention 
mainly to the measurement of the velocity dispersion of the cluster in 
NGC~1569 and discussion of its implications.  Ho \& Filippenko (1996) 
present data for additional similar star clusters, and a future paper will 
analyze other information on the stellar population, gas kinematics, and 
interstellar absorption.

\section{Observations and Analysis}

On 9 January 1996 UT, we took high-dispersion spectra of 
the brighter of the two prominent clusters in NGC~1569 (cluster ``A'', 
hereafter NGC~1569-A; $V$ = 14.8 mag; O'Connell \etal 1994) using the HIRES 
echelle spectrograph (Vogt \etal 1994) with the Keck 10~m telescope on Mauna 
Kea, Hawaii.  Four consecutive half-hour exposures were obtained through a 
1\farcs15 $\times$ 7\asec\ slit.  The full spectral range recorded 
wavelengths from $\sim$3900 \AA\ to 6280 \AA\ in 34 spectral orders.  The final 
spectral resolution, as determined from the profiles of the comparison lamp 
lines, is $R\,\approx$ 38,000 (full width at half maximum = 7.9 \kms).
We also took brief exposures of several bright stars of known spectral types 
for application of the cross-correlation method to derive velocity dispersions.
The program object and standard stars were interleaved with exposures of 
thorium and argon hollow cathode comparison lamps to monitor shifts in the 
wavelength scale.  The initial data reduction closely followed standard 
procedures for echelle spectroscopy (e.g., Ho \& Filippenko 1995), with a 
few minor modifications described in Ho \& Filippenko (1996), where additional 
details concerning the observing strategy and instrument setup are also given.
The one-dimensional spectra were extracted with a constant effective 
aperture of 1\farcs15 $\times$ 2\farcs05.  The background signal was 
determined by averaging two adjacent regions on either side of the object. 
Within the extraction aperture, the light from the underlying galaxy is about 
5 times fainter than that from the cluster; moreover, the stellar features 
of the background are very weak, and inaccuracies in background subtraction 
should have a minor effect on our analysis.

Our principal aim is to derive the line-of-sight velocity dispersion of the
stellar component of NGC~1569-A.  A number of studies based on different lines 
of evidence have concluded that the cluster is in its ``post-burst'' phase, 
with an age of $\sim$10--20 Myr (Israel 1988; Israel \& de Bruyn 
1988; Waller 1991; O'Connell \etal 1994).  The relative youth of the cluster 
poses a set of complications not normally encountered in the measurement
of velocity dispersions of old stellar populations.  Based on a blue 
($\sim$3200--4500 \AA) spectrum of NGC~1569-A, Arp \& Sandage (1985) determined 
an average spectral type of A0~Iab for the cluster.  While weak metal lines 
can be easily discerned in our high-dispersion spectra at these wavelengths, 
they cannot be used for velocity dispersion measurement in our case because 
the line widths expected are dominated by other sources of line 
broadening intrinsic to stars of this spectral type.  In early-A supergiants, 
both macroturbulent and microturbulent motions in the unstable atmospheres, as 
well as mild rotational broadening, broaden the lines by an amount 
comparable to the velocities anticipated from the virial motions of the 
individual stars. 
%(e.g., Boer, de Jager, \& Nieuwenhuijzen 1988).
Hence, one must use portions of the spectrum whose flux is not dominated by 
early-type supergiants.

We base our analysis on the region of the spectrum from $\sim$5000 to 6280 
\AA, which we argue below is suitable for measuring velocity dispersions.  
According to stellar population models (e.g., Bruzual \& Charlot 1993), the 
spectrum of a cluster with an age of $\sim$10 Myr comes largely from 
supergiants at wavelengths near and redward of the $V$ band.  In particular, a 
substantial fraction of the light should come from cool (F--M) supergiants 
(O'Connell 1996), as supported by the detection of strong Ca~II infrared 
triplet absorption lines (Prada \etal 1994).  Although the contribution from 
early (A and B) supergiants at these wavelengths is still not negligible, their 
spectra are relatively featureless compared to those of cool supergiants 
(e.g., Jacoby, Hunter, \& Christian 1984).  We neglected to take spectra of 
early-type supergiants during our observing run, but J.~K. McCarthy kindly 
provided us with a spectrum of the star 177-A, a low-metallicity A0 supergiant 
in M33 observed with HIRES in a configuration nearly identical to the one used 
here (McCarthy \etal 1995).  The comparison with the M33 star may be 
appropriate given the low oxygen abundance of NGC~1569 (Hunter, Gallagher, \& 
Rautenkranz 1982).  We confirmed that very few metal lines are found 
in the A0~Ia star at these wavelengths, and virtually none of the lines seen 
in the spectrum of NGC~1569-A are visible.  By contrast, nearly
all of the features seen in the cluster spectrum can be identified with metal 
lines in the G to M giants and supergiants we observed.  Similarly, the flux of 
a 10-Myr old cluster at these wavelengths also comes partly from early-B stars 
still on the main sequence (O'Connell 1996), but such stars were verified to be 
nearly featureless in the region of interest.  

To what extent can velocity dispersions be measured from an integrated 
spectrum dominated by the light of cool supergiants?  In studies of old 
stellar populations (e.g., Illingworth 1976; Tonry \& Davis 1979),
the velocity template stars used are red {\it giants}, whose lines are 
intrinsically narrow and generally unresolved.  Due principally to the effects 
of macroturbulence, the line widths of cool supergiants, on the other hand, 
are {\it not} negligible, typically having $\sigma\,\approx$ 9 \kms\ for types 
F5--K5, with a spread about the mean of perhaps 1.5 \kms\ (Gray \& Toner 
1987).  However, as long as the mass of the cluster is not too small, we 
should be able to extract its velocity dispersion using a cool supergiant as 
template.

The pattern of absorption lines in NGC~1569-A grossly resembles that of the
template HR~2289 (46 $\psi^{\prime}$ Aur; see Ho \& Filippenko 1996), listed 
as K5--M0 Iab-Ib in the Bright Star Catalog, although in detail it seems to 
match that of HR~3422 (G8~IV) more closely, suggesting that the cluster 
supergiants may have somewhat higher effective temperatures than K5--M0.  
Unfortunately, HR~2289 was the only supergiant we observed.  To obtain a 
preliminary result, we adopt the following strategy. We use HR~3422 as the 
template to obtain an initial value of the velocity dispersion, and then we 
subtract from it in quadrature an amount expected to be due to the intrinsic 
widths of the supergiants.  Note that using a subgiant as the template 
should not affect the derived dispersion, since rotational broadening 
in cool subgiants remains insignificant (Smith \& Dominy 1979) and their
macroturbulent motions are even smaller than in late-type giants 
(Gray \& Toner 1986).

We applied the cross-correlation method of Tonry \& Davis (1979) to 
the spectral orders between 5000 \AA\ and 6280 \AA, using the G8 IV star
as the template.  After excluding several orders having low signal-to-noise 
ratios (S/N) and/or lines of very low contrast, the final 6 usable orders 
yielded an average velocity dispersion of 18.1 \kms, with a standard deviation 
of 1.1 \kms.  An example of one of the orders is shown in Figure 1.
Assuming that cool supergiants have intrinsic line 
widths of 9 \kms\ (Gray \& Toner 1987), we estimate that the line-of-sight
component of the velocity dispersion arising from gravitational 
motions is $\sigma_*$ = 15.7$\pm$1.5 \kms, where the error is simply a 
conservative guess of the uncertainty in our procedure.  This value
lies near the top end of the distribution of velocity dispersions for Galactic 
globular clusters (Illingworth 1976; Mandushev, Spassova, \& Staneva 1991).

\section{The Dynamical Mass of the Cluster}

If NGC~1569-A is gravitationally bound, the virial theorem can be 
used to obtain the total mass of the cluster, provided that an effective 
gravitational radius can be measured for the system: $M\,=\,3\sigma_*^2 R/G$.  
This simple relation assumes that (1) all the stars have equal 
masses, (2) the cluster is spherically symmetric, and (3)
the velocity distribution is isotropic [$\sigma^2$(total) = 3$\sigma_*^2$].
Unlike the case of star clusters in the Galaxy and in some members of the 
Local Group, a detailed radial profile is not yet available for NGC~1569-A, 
and hence we cannot apply the somewhat more sophisticated formalism  
described, for  instance, by Illingworth (1976).  
%%The observations of 
%NGC~1569-A presented in O'Connell \etal (1994) were acquired prior to the 
%refurbishment mission of {\it HST}; consequently, the radial profile of 
%the cluster was severely affected by the aberrated optics. O'Connell \etal 
%quote a half-light radius of $R_{\rm h}\,<$ 2.2 pc (assuming a distance of 
%2.5 Mpc).  Applying a different method of analysis to the same data, Meurer 
%\etal (1995) obtained $R_{\rm h}$ = 1.9$\pm$0.2 pc (adjusted to the distance 
%preferred by O'Connell  et al.).  If we adopt the size found by Meurer \etal 
%and assume that $R_{\rm h}$ gives an approximate measure of the effective 
%radius used in the virial equation (see Ho \& Filippenko 1996), we obtain a 
%mass of $M$ = (3.3$\pm$0.5)\e{5} \solmass.  
Adopting a half-light radius of 1.9$\pm$0.2 pc (from Meurer \etal 
1995, after adjusting to the distance of 2.5 Mpc preferred by 
O'Connell et al. 1994) as a reasonable approximation of the effective 
radius, we obtain $M$ = (3.3$\pm$0.5)\e{5} \solmass.  

How reliable is this mass estimate?  Let us briefly examine the likely 
consequences of our main assumptions.  The use of the virial theorem to 
derive the mass requires the cluster to be bound and dynamically relaxed.  
The first condition seems very likely to be satisfied; given the compactness 
of the cluster, any plausible expansion timescale is shorter than 1 Myr,
whereas the estimated age of the cluster is at least an order of magnitude 
larger.  However, if NGC~1569-A is indeed as massive as a typical 
globular cluster, it is unlikely to have completely virialized, since 
the relaxation time of most globular clusters is on the order of 10$^9$ 
yr (Binney \& Tremaine 1987).  Our measured velocity dispersion therefore 
underestimates the value it will attain when the cluster is truly virialized, 
and the derived mass represents a lower limit.  The assumption that all 
the stars have equal masses obviously is a gross oversimplification.  However, 
relaxing it will also increase our mass estimate.  From a comparison of 
globular clusters whose dynamical masses have been determined using both 
single-mass and multi-mass models, Mandushev \etal (1991) conclude that the 
former tends to underestimate the masses by about a factor of two.  Finally, 
from a comparison of velocity dispersions obtained from spatially integrated 
spectra versus those computed from radial velocities of individual stars, 
Zaggia, Capaccioli, \& Piotto (1993) find that the former method 
systematically biases the dispersions to lower values.

Of all the conventional parameterizations of the cluster size, the half-light 
radius appears to be the most robust characterization, since it is the 
least sensitive to evolutionary or environmental effects (van den Bergh, 
Morbey, \& Pazder 1991).  The angular size of NGC~1569-A, however, is not 
known to high accuracy (the images were taken prior to the {\it HST} 
refurbishment mission), and lack of a reliable distance determination to the 
galaxy further blurs its true linear size.  Our adopted distance of 2.5 Mpc 
stems from a rather provisional evaluation of the resolved supergiant 
population surrounding the cluster (O'Connell \etal 1994).  A survey 
of other published distances for the galaxy (e.g., Israel 1988; Tully 1988; 
Hunter \etal 1989) gives a range of 1.6 to 4.7 Mpc.
Nevertheless, since the size of the cluster only enters linearly into the 
virial equation, its uncertainty has a less severe effect on the mass estimate 
than that associated with the velocity dispersion.

Previous mass estimates of NGC~1569-A and other objects of its kind (see 
\S\ 1) have been highly uncertain, since they invariably 
relied on stellar population models that depend on a large 
number of poorly constrained parameters.  The total stellar mass, in 
particular, is very difficult to obtain, since virtually all of the observables 
in such clusters trace the young, massive stars, which comprise only a fraction
of the total mass for a normal initial mass function.  The present study aims 
to bypass these complications by obtaining a direct measurement of the 
dynamical mass.  Although the exact mass of NGC~1569-A is still difficult to 
pin down at the moment, there seems to be little doubt that it is indeed quite 
large, most likely on the order of (2--6)\e{5} \solmass.  This value falls 
comfortably within the range of masses of Galactic globular clusters, whose 
average and median values are, respectively, 1.9\e{5} and 8.1\e{4} \solmass\ 
(Mandushev \etal 1991).  Most of the assumptions inherent in our calculation, 
in fact, bias the mass toward lower values; hence, the true mass may be even 
higher than our nominal estimate of 3.3\e{5} \solmass.  NGC~1569-A is extremely 
compact: its half-light radius 
is merely 1.9 pc, comparable to, and perhaps a bit smaller than, that of the 
average globular cluster (Mandushev \etal 1991; van den Bergh \etal 1991).  It 
follows, therefore, that its mass density (1.1\e{4} \solmass\ pc$^{-3}$) is at 
least as large as that of typical Galactic globular clusters.  Since 
dynamical evolution will cause the cluster to expand as it ages (Elson 1992), 
the density of NGC~1569-A may not be very unusual.  

Finally, it is worth considering the mass-to-light ratio 
of the cluster.  With standard parameters, evolutionary synthesis calculations 
(e.g., Bruzual \& Charlot 1993) predict that the visual light of a 10-Myr 
cluster will fade by about 6--7 mag in 10--15 Gyr.  With $M_V$ = 
--14.1 mag currently (O'Connell \etal 1994), the evolved cluster will dim to 
$M_V$ = --7 to --8, again lying comfortably within the peak of the luminosity 
function of globular cluster systems ($\langle M_V \rangle\,\approx$ --7.3 
mag; Harris 1991).  NGC~1569-A will have $M/L_V$ = 2.5--6.3 
($M/L_V$)$_{\odot}$, 
comparable to and perhaps slightly larger than normal for Galactic 
globular clusters [0.7--2.9 ($M/L_V$)$_{\odot}$; Mandushev \etal 1991].  
This finding implies that, to a first approximation, the stellar initial 
mass function of NGC~1569-A is similar to that of typical globular clusters.

\section{Conclusions}

We obtained high-dispersion optical spectra of one of the two luminous 
compact clusters (object ``A'') located in the central region of the dwarf 
galaxy NGC~1569. Cool supergiants contribute significantly to the light at 
visual wavelengths, and we argue that it is possible to measure the velocity 
dispersion of the cluster using a conventional cross-correlation technique.  
The velocity dispersion along the line of sight is estimated to be $\sigma_*$ =
15.7$\pm$1.5 \kms.  Combined with the size measurement known from 
{\it HST} images, a dynamical mass of $M$ = (3.3$\pm$0.5)\e{5} \solmass\ 
is determined using the virial theorem.  The derived mass, mass density, and 
probable mass-to-light ratio of NGC~1569-A provide compelling evidence that 
the cluster will evolve into a fairly massive globular cluster.  
%Moreover, the probable mass-to-light ratio of the cluster after aging 
%10--15 Gyr [2.5--6.3 ($M/L_V$)$_{\odot}$] also seems to be roughly compatible 
%with the range seen in Galactic globular clusters.  
Further observations of this kind are needed to establish whether 
other luminous, compact, young star clusters being discovered in 
starburst environments are similar in nature to NGC~1569-A.

\acknowledgments

The W. M. Keck Observatory, made possible by the generous and visionary 
gift of the W.~M. Keck Foundation, is operated as a scientific partnership 
between the California Institute of Technology and the University of 
California.  We thank Tom Bida for his proficient guidance on the use of 
HIRES, Meg Whittle and Joel Aycock for their technical support, and Aaron 
Barth for help in planning some of the observations.  We are grateful to John 
Stauffer and to the referee for advice concerning analysis of the spectra, to 
James McCarthy for sending his spectrum of 177-A, and to Lewis Jones and Bob 
O'Connell for pertinent discussions on stellar populations.  The research of 
L.~C.~H. is funded by a  postdoctoral fellowship from the Harvard-Smithsonian 
Center for Astrophysics, while A.~V.~F. receives financial support from the 
National Science Foundation (grant AST-9417213) and NASA (grant 
AR-05792.01-94A from the Space Telescope Science Institute).  Partial travel 
support was provided by the California Association for Research in Astronomy.  
During the course of this work, A.~V.~F. held an appointment as a Miller 
Research Professor in the Miller Institute for Basic Research in Science 
(U.~C. Berkeley).  

%\clearpage
%\appendix

%REFERENCES
\clearpage

\centerline{\bf{References}}
\medskip
%\begin{references}

\refindent 
Arp, H., \& Sandage, A. 1985, AJ, 90, 1163

\refindent 
Barth, A.~J., Ho, L.~C., Filippenko, A.~V., \& Sargent, W.~L.~W. 1995, \aj,
110, 1009

\refindent 
Benedict, G.~F., \etal 1993, \aj, 105, 1369

\refindent 
Binney, J., \& Tremaine, S. 1987, Galactic Dynamics (Princeton: Princeton 
Univ. Press)

%\refindent 
%Boer, B., de Jager, C., \& Nieuwenhuijzen, H. 1988, \aa, 195, 218

\refindent 
Bower, G.~A., \& Wilson, A.~S. 1995, \apjs, 99, 543

\refindent 
Bruzual A., G., \& Charlot, S. 1993, \apj, 405, 538

%\refindent 
%Charlot, S., \& Bruzual A., G. 1991, \apj, 367, 126

\refindent 
Conti, P.~S., Leitherer, C., \& Vacca, W.~D. 1996, \apj, 461, L87

\refindent 
Conti, P.~S., \& Vacca, W.~D. 1994, \apj, 423, L97

%\refindent 
%de Koter, A., de Jager, C., \& Nieuwenhuijzen, H. 1988, \aa, 200, 146

%\refindent 
%Drissen, L., \& Roy, J.-R. 1994, \pasp, 106, 974
%
%\refindent 
%Drissen, L., Moffat, A.~F.~J., \& Shara, M.~M. 1993, \aj, 105, 1400

%\refindent 
%Drissen, L., Moffat, A.~F.~J., Walborn, N.~R., \& Shara, M.~M. 1995, \aj, 110,
%2235

%\refindent 
%Drissen, L., Roy, J.-R., \& Moffat, A.~F.~J. 1993, \aj, 106, 1460

%\refindent 
%Dubath, P., Mayor, M., \& Meylan, G. 1993, in The Globular Cluster-Galaxy
%Connection, ed. G.~H. Smith \& J.~P. Brodie (San Francisco: ASP), 557

\refindent 
Elson, R.~A.~W. 1992, \mnras, 256, 515

%\refindent 
%Filippenko, A.~V. 1982, \pasp, 94, 71

%\refindent 
%Friel, E.~D. 1993, in The Globular Cluster-Galaxy Connection, ed. G.~H. Smith
%\& J.~P. Brodie (San Francisco: ASP), 273

\refindent 
Gray, D.~F., \& Toner, C.~G. 1986, \apj, 310, 277

\refindent 
Gray, D.~F., \& Toner, C.~G. 1987, \apj, 322, 360

\refindent 
Harris, W.~E. 1991, \annrev, 29, 543

%\refindent 
%Heckman, T.~M., Dahlem, M., Lehnert, M.~D., Fabbiano, G., Gilmore, D., \&
%Waller, W.~H. 1995, \apj, 448, 98

%\refindent 
%Ho, L.~C., \etal 1996, in preparation

\refindent 
Ho, L.~C., \& Filippenko, A.~V. 1995, \apj, 444, 165

\refindent 
Ho, L.~C., \& Filippenko, A.~V. 1996, \apj, submitted

\refindent 
Holtzman, J.~A., \etal 1992, \aj, 103, 691

%\refindent 
%Horne, K. 1986, \pasp, 98, 609

\refindent 
Hunter, D.~A., Gallagher, III, J.~S., \& Rautenkranz, D. 1982, \apjs, 49, 53

\refindent 
Hunter, D.~A., O'Connell, R.~W., \& Gallagher, III, J.~S. 1994, \aj, 108, 84

\refindent 
Hunter, D.~A., Thronson, H.~A., Jr., Casey, S., \& Harper, D.~A. 1989, \apj,
341, 697

\refindent 
Illingworth, G. 1976, \apj, 204, 73

\refindent 
Israel, F.~P. 1988, \aa, 194, 24

\refindent 
Israel, F.~P., \& de Bruyn, A.~G. 1988, \aa, 198, 109

\refindent 
Jacoby, G.~H., Hunter, D.~A., \& Christian, C.~A. 1984, \apjs, 56, 257

%\refindent 
%Leitherer, C., \& Heckman, T.~M. 1995, \apjs, 96, 9

\refindent 
Leitherer, C., Vacca, W.~D., Conti, P.~S., Filippenko, A.~V., Robert, C., \&
Sargent, W.~L.~W. 1996, \apj, in press

%\refindent 
%Lennon, D.~J., Dufton, P.~L., \& Fitzsimmons, A. 1992, \aas, 94, 569

%\refindent 
%Lupton, R.~H., Fall, S.~M., Freeman, K.~C., \& Elson, R.~A.~W. 1989, \apj,
%347, 201

\refindent 
Lutz, D. 1991, \aa, 245, 31

\refindent 
Mandushev, G., Spassova, N., \& Staneva, A. 1991, \aa, 252, 94

\refindent 
Maoz, D., Barth, A.~J., Sternberg, A., Filippenko, A.~V.,
Ho, L.~C., Macchetto, F.~D., Rix, H.-W., \& Schneider, D.~P. 1996, \aj,
in press

\refindent 
McCarthy, J.~K., Lennon, D.~J., Venn, K.~A., Kudritzki, R.-P., Puls, J., \&
Najarro, F. 1995, \apj, 455, L135

\refindent 
Melnick, J., Moles, M., \& Terlevich, R. 1985, \aa, 149, L24

%\refindent 
%Meurer, G.~R., Freeman, K.~C., Dopita, M.~A., \& Cacciari, C. 1992, \aj, 103,
%60

\refindent 
Meurer, G.~R., Heckman, T.~M., Leitherer, C., Kinney, A., Robert, C., \&
Garnett, D.~R. 1995, \aj, 110, 2665

\refindent 
O'Connell, R.~W. 1996, private communication

\refindent 
O'Connell, R.~W., Gallagher, J.~S., \& Hunter, D.~A. 1994, \apj, 433, 65

\refindent 
O'Connell, R.~W., Gallagher, J.~S., Hunter, D.~A., \& Colley, W.~N. 1995, \apj,
446, L1

%\refindent 
%Oke, J.~B., \& Gunn, J.~E. 1983, \apj, 266, 713

\refindent 
Prada, F., Greve, A., \& McKeith.~D. 1994, \aa, 288, 396

%\refindent 
%Sargent, W.~L.~W., \& Filippenko, A.~V. 1991, \aj, 102, 107

%\refindent 
%Sargent, W.~L.~W., Schechter, P.~L., Boksenberg, A., \& Shortridge, K. 1977,
%\apj, 212, 326

\refindent 
Schweizer, F., \& Seitzer, P. 1993, \apj, 417, L29

\refindent 
Smith, M.~A., \& Dominy, J.~F. 1979, \apj, 231, 477

\refindent 
Tonry, J., \& Davis, M. 1979, \aj, 84, 1511

\refindent 
Tully, R.~B. 1988, Nearby Galaxies Catalog (Cambridge: Cambridge Univ. Press)

\refindent 
Vacca, W.~D. 1994, in Violent Star Formation, ed. G. Tenorio-Tagle
(Cambridge Univ. Press), 297

\refindent 
van den Bergh, S., Morbey, C., \& Pazder, J. 1991, \apj, 375, 594

\refindent 
Vogt, S.~S., \etal 1994, Proc. SPIE, 2198, 362

\refindent 
Waller, W.~H. 1991, \apj, 370, 144

\refindent 
Whitmore, B.~C., \& Schweizer, F. 1995, \aj, 109, 960

\refindent 
Whitmore, B.~C., Schweizer, F., Leitherer, C., Borne, K., \& Robert, C. 1993,
\aj, 106, 1354

\refindent 
Zaggia, S.~R., Capaccioli, M., \& Piotto, G. 1993, \aa, 278, 415

\refindent 
Zepf, S.~E., Carter, D., Sharples, R.~M., \& Ashman, K.~M. 1995, \apj, 445,
L19

%\end{references}

%FIGURE CAPTIONS
\clearpage

\centerline{\bf{Figure Captions}}
\medskip
Fig. 1. --- Example of the cross-correlation technique applied to one of the
spectral orders.  The {\it top} panel shows the spectrum of NGC~1569-A
and the {\it middle} panel the template star HR~3422 (G8 IV), both normalized 
to unity and shifted to their rest frame.  The cluster spectrum has been 
smoothed with a boxcar function of 3 pixels (6.4 \kms) in order to slightly 
improve the S/N for the sake of the presentation.  The {\it bottom} panel 
plots the cross-correlation function (CCF) between the cluster and the star.  
%The velocity of the main peak of the CCF corresponds to the relative 
%velocity between the object and the template (the peak here does not give the 
%absolute radial velocity of NGC~1569-A since the template is not at zero
%velocity).  
The width of the main peak of the CCF is related to the velocity dispersion 
of the object, and the relation between the quantities is determined by 
empirical tests.  In this order, the S/N per pixel of the continuum in the 
cluster (before smoothing) ranges from 80 at the blue end to 120 at the red 
end.  The corresponding values for the star are S/N = 320 to 490.

%FIGURES
\clearpage
\begin{figure}
\plotone{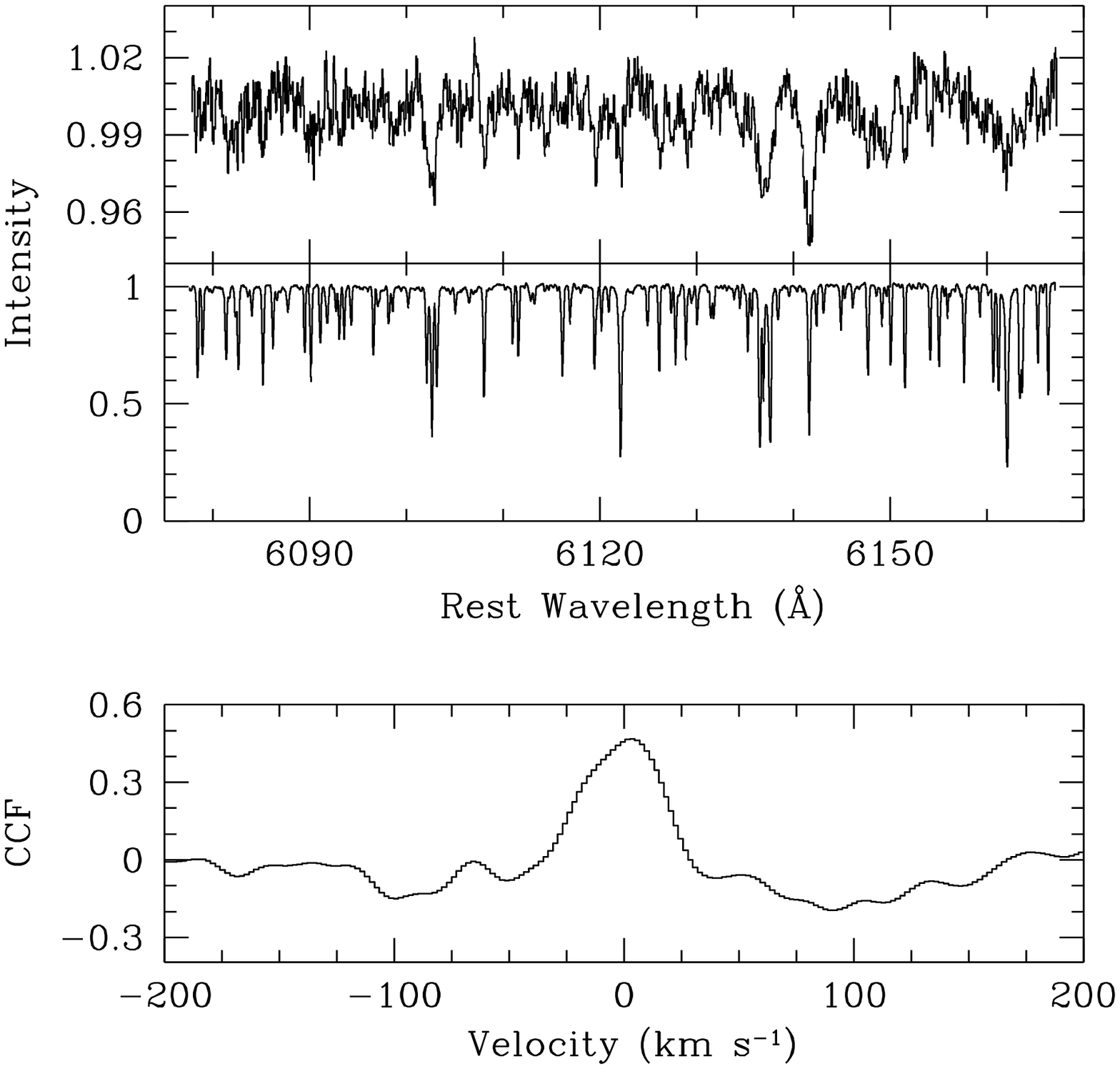}
\caption{Figure 1.}
\end{figure}

%\clearpage
%\begin{figure}
%\plotone{fig1b.ps}
%\caption{Figure 1{\it b}.}
%\end{figure}

\end{document}